# Noise minimization in optical detection of small particles

Taras Plakhotnik

School of Physical Sciences, University of Queensland, Brisbane, QLD 4072, Australia

## Abstract

The ultimate sensitivity of optical detection is limited by the signal-to-noise ratio (SNR). The first part of the paper shows that coherence plays an important role in the noise analysis. Although interference between an auxiliary wave and a signal wave makes the photo detector response to the signal stronger, the coherent noise also enhances. This makes insignificant the gain in the SNR. Pulsed-excitation gated-detection (PEGD) is described and analyzed in the second part to show that 1) a high brightness of detected particles is not a prerequisite for a high SNR, 2) optimized parameters of the PEGD protocol demonstrate interesting bifurcation making a sudden jump from an effectively continuous regime to PEGD, and 3) photo-physical properties of $NV^-$ centers in nano crystals of diamond approach those ideal for PEGD.

**Introduction**

Imagine that you can follow the course of a single virus for several hours and analyse its infection path [1] or that you are able to monitor delivery of few drag molecules to a single cell. This breath taking perspective may become one day a reality. The challenge is to signal a specific process out of several others running simultaneously on crowded background. A promising idea for selective visualisation of physical and chemical dynamics in biological and other complex systems is based on labelling the objects of interest with chemically inert, nanometre-sized, markers which can be non-invasively and infinitely long observed using appropriate optical techniques [1, 2, 3]. Although the selectivity depends on the contrast which the marker has against the background and a higher contrast is an advantage, the ultimate limitation for detection is a signal-to-noise ratio (SNR). This paper examines ways for noise minimisation in optical detection of nanoparticles.

Generally, all markers can be divided into two groups – depending whether coherent or incoherent emission is the dominating signal sent by the particle to an observer or a photo detector. Direct detection of absorption is, as a rule, more difficult [4] and will not be considered in this paper.

A coherent wave can be manipulated and analyzed using interferometers and other setups where interference plays a role. For example, a scheme has been proposed for single-molecule detection by Plakhotnik et al [5], where the scattered wave interferes on the detector with an auxiliary reference wave. This idea has been explored by Lindfors et al. [6] and by Ignatovich et al [7] for gold and other particles and has been advertised with controversial claims about its superior SNR. We will analyse this type of measurements using a general description to show that, as a rule, only incoherent noise can be eliminated in any interferometric scheme.

Incoherent emission (usually luminescence) has an advantage of having specific spectral and temporal characteristics. Such specificity can be exploited for SNR enhancement. This paper is focused on optimizing the temporal response. We will see that high brightness (number of photons emitted per unit time) of a marker is not necessarily synergetic with a high SNR in the detected signal.

**Coherent signal**

When several fields overlap on a photo detector, the photo current reads

$$J(t) \propto \int_{-\infty}^{t} g(t,t') \int_{A} \theta(\mathbf{r}) |\mathbf{E}_b(\mathbf{r},t') + \mathbf{E}_s(\mathbf{r},t') + \mathbf{E}_r(\mathbf{r},t')|^2 \, dAdt' \qquad (1)$$

where $g(t,t')$ describes the time response of the photo detector. Function $\theta(\mathbf{r})$ is equal to 1 for ordinary detectors but if a split-detector is used, then $\theta(\mathbf{r})=1$ on the one half of the detector and $\theta(\mathbf{r})=-1$ on the other half. The total intensity is integrated over the detector area $A$. The split-detector removes contribution from space symmetric fluctuations, for example, fluctuations of power [7]. The fields relevant to the problem of detecting a nanoparticle are 1) the field scattered by the particle $\mathbf{E}_s(\mathbf{r},t')$ expressed in a complex representation, 2) the reference field $\mathbf{E}_r(\mathbf{r},t')$, and 3) the background field $\mathbf{E}_b(\mathbf{r},t')$. The background field comes from the sample and obscures the scattered field. In Eq. (1) the fields are treated classically and the noise is taken into account by assuming that every field fluctuates. When the overall noise of the acquired data is calculated, $\mathbf{E}_s(\mathbf{r},t')$ in (1) can be neglected because the scattered field is relatively week. The dominating noise is related to the background field which can be written in the form $\mathbf{E}_b \equiv \left[\overline{\mathbf{E}}_b + \boldsymbol{\varepsilon}\right]\exp(-i\omega t)$, where $\boldsymbol{\varepsilon}$ represents noise and the bar on the top indicates averaging the complex amplitude over time. For briefness, we will also use a shorter notation $\langle \mathbf{F} \rangle_A \equiv A^{-1} \int_A \theta(\mathbf{r}) \mathbf{F}(\mathbf{r},t) dA$ for the detector area averaging and $\langle \mathbf{F} \rangle_{TR} \equiv \int_{-\infty}^{t} g(t,t') \mathbf{F} dt'$ for averaging of any function $\mathbf{F}$ over the detector time-response. Without narrowing too much the applicability of the following calculations, we will make a reasonable assumption that the probability distribution of $\boldsymbol{\varepsilon}(\mathbf{r},t)$ is symmetric with respect to zero at every point $\mathbf{r}$ of the detector. This assumption allows to disregard all terms proportional to an odd power of $\boldsymbol{\varepsilon}$. The question is – under what conditions can the addition of the reference field $\mathbf{E}_r(\mathbf{r},t')$ improve the SNR?

The variance of the photo current is given by

$$\text{var}[J_p] \propto \overline{\left\{2\,\text{Re}\left\langle \left[\left(\overline{\mathbf{E}}_r^* + \overline{\mathbf{E}}_b^*\right)\langle \boldsymbol{\varepsilon} \rangle_{TR}\right]\right\rangle_A\right\}^2} + \text{var}\left[\langle |\boldsymbol{\varepsilon}|^2 \rangle_{TRA}\right], \qquad (2)$$

and the corresponding SNR reads

$$SNR = \frac{2\left|\overline{\mathbf{E}}_r + \overline{\mathbf{E}}_b\right| \left\langle \left|\langle \mathbf{E}_s \rangle_{TR}\right| \cdot \cos\phi_s \right\rangle_A + \left\langle |\mathbf{E}_s|^2 \right\rangle_{TRA}}{\left\{4\left|\overline{\mathbf{E}}_r + \overline{\mathbf{E}}_b\right|^2 \overline{\left\langle \left|\langle \boldsymbol{\varepsilon} \rangle_{TR}\right| \cdot \cos\phi_\varepsilon \right\rangle_A^2} + \text{var}\left[\langle |\boldsymbol{\varepsilon}|^2 \rangle_{TRA}\right] + \text{var}[D_e]\right\}^{1/2}}, \qquad (3)$$

where we have included a variance $\text{var}[D_e]$ of the detector output caused by electrical noise and have assumed for simplicity that $\overline{\mathbf{E}}_r$ and $\overline{\mathbf{E}}_b$ are position independent. The values of $\phi_s$ and $\phi_\varepsilon$ are the phase differences between $\overline{\mathbf{E}}_r + \overline{\mathbf{E}}_b$ and the signal and noise fields respectively. Under special circumstances, for example, when $\phi_\varepsilon = \pi/2$ at every point of the detector area (this is possible only if the intensity but not the phase of the background fluctuates), the SNR is proportional to $\overline{\mathbf{E}_r}$. Note that $\phi_s$ should not be equal to $\pi/2$ for this to happen.

In a general case, the reference field will enhance both the signal and the noise. It follows from Eq. (3) that the maximum achievable SNR satisfies the relation

$$SNR_{max}^2 = SNR_0^2 + \frac{\left\langle \left|\langle \mathbf{E}_s \rangle_{TR}\right| \cdot \cos\phi_s \right\rangle_A^2}{\left\langle \left|\langle \boldsymbol{\varepsilon} \rangle_{TR}\right| \cdot \cos\phi_\varepsilon \right\rangle_A^2} \leq SNR_0^2 + SNR_0 \frac{\left\{ \mathrm{var}\left[\left\langle |\boldsymbol{\varepsilon}|^2 \right\rangle_{TRA}\right] + \mathrm{var}[D_e] \right\}^{1/2}}{\left\langle \left|\langle \boldsymbol{\varepsilon} \rangle_{TR}\right| \cdot \cos\phi_\varepsilon \right\rangle_A^2}, \quad (4)$$

where $SNR_0$ is the SNR when $\left|\overline{\mathbf{E}_r} + \overline{\mathbf{E}_b}\right| = 0$ and the estimate is obtained using the Schwartz inequality. Reliable detection of a particle requires $SNR_{max} \gg 1$ while the condition $SNR_{max} \gg SNR_0$ justifies the complications associated with the addition of the auxiliary reference beam to the setup. For satisfying these two inequalities, the relation

$$\left\{ \mathrm{var}\left[\left\langle |\boldsymbol{\varepsilon}|^2 \right\rangle_{TRA}\right] + \mathrm{var}[D_e] \right\}^{1/2} \gg \overline{\left\langle \left|\langle \boldsymbol{\varepsilon} \rangle_{TR}\right| \cdot \cos\phi_\varepsilon \right\rangle_A^2} \quad (5)$$

must hold. This can be seen from (4) by considering consecutively cases of $SNR_0 > 1$ and $SNR_0 < 1$. If the electrical noise is negligible, the condition for substantial improvement of the SNR is $\mathrm{var}\left[\left\langle |\boldsymbol{\varepsilon}|^2 \right\rangle_{TRA}\right]^{1/2} \gg \overline{\left\langle \left|\langle \boldsymbol{\varepsilon} \rangle_{TR}\right| \cdot \cos\phi_\varepsilon \right\rangle_A^2}$. Generally, the probability of very large spikes in the photocurrent noise can be significant, but for the most typical

Gaussian noise the variance on the left side is close to $\left(\overline{\left\langle |\boldsymbol{\varepsilon}|^2 \right\rangle_{TRA}}\right)^2$. In such a case the latest inequality is equivalent to the condition of temporal/spatial incoherence of the background noise $\overline{\left\langle |\boldsymbol{\varepsilon}|^2 \right\rangle_{TRA}} \gg \overline{\left\langle |\left\langle \boldsymbol{\varepsilon} \right\rangle_{TR}| \cdot \cos\phi_\varepsilon \right\rangle_A^2}$. This relation and an obvious requirement $\left\langle |\left\langle \mathbf{E}_s \right\rangle_{TR}| \cdot \cos\phi_s \right\rangle_A^2 \approx \left\langle |\mathbf{E}_s|^2 \right\rangle_{TRA}$ that the cross-term between $\overline{\mathbf{E}}_r + \overline{\mathbf{E}}_b$ and $\mathbf{E}_s$ is not being washed out by the averaging are the conditions for background noise suppression by a stable auxiliary wave.

**Incoherent signal**

When the response of the marker is incoherent, temporal characteristics of this response can be used to improve the SNR. The pulsed-excitation gated-detection (PEGD) protocol works as follows. A pulsed laser periodically excites emission of the marker which is then integrated for time $\tau_i$. Every exciting pulse is assumed to be very short ($\delta$-pulse) and the integration starts with a delay $\tau_d$ after each pulse. The laser pulse excites both emission of the marker characterized with a decay time $\tau_s$ and a background decaying with a time constant $\tau_b$. A continuous excitation/detection (CECD) scheme is obtained by taking the limit $\tau_d \to 0$ and $\tau_i \to 0$. The transition from PEGD to CECD is gradual but it will be shown that an almost discontinuous jump from $\tau_d, \tau_i \gg \tau_b$ to $\tau_d, \tau_i = 0$ happens when the measurement protocol is optimized towards the highest SNR for different values of $\tau_s$.

The total number of signal photons detected within the measuring time $\tau$ is $n_s \tau/(\tau_d + \tau_i)$, where the number of photons detected in a single pulse is

$$n_s = \int_{\tau_d}^{\tau_d+\tau_i} \frac{a_s}{\tau_s}\exp\left(-\frac{t}{\tau_s}\right)dt = a_s\left[\exp\left(-\frac{\tau_d}{\tau_s}\right) - \exp\left(-\frac{\tau_d+\tau_i}{\tau_s}\right)\right]. \qquad (6)$$

The total number of detected background photons is $n_b \tau/(\tau_d + \tau_i)$, where $n_b$ is given by Eq. (6) when all subscripts $s$ are substituted with $b$. The total number of photons in the background and signal pulses are $a_b$ and $a_s$ respectively. If the signal is generated by a single quantum emitter, then $a_s < 1$ since a single emitter can emit only one photon per pulse. The SNR in such gated measurements reads

$$SNR_p = \frac{n_s}{\left[n_s + n_b + (n_s+n_b)^2 \upsilon/\tau_i\right]^{1/2}}\left(\frac{\tau}{\tau_i+\tau_d}\right)^{1/2} \qquad (7)$$

The denominator in Eq. (7) includes two contributions. The variance of the Poissonian shot noise is given by $n_s + n_b$. The remaining part under the square root accounts for power noise. In the CECD limit, the SNR equals

$$SNR_c = \frac{a_s \dfrac{\tau}{\tau_s}}{\left[a_s\dfrac{\tau}{\tau_s} + a_b\dfrac{\tau}{\tau_b} + \dfrac{\upsilon}{\tau}\left(a_s\dfrac{\tau}{\tau_s} + a_b\dfrac{\tau}{\tau_b}\right)^2\right]^{1/2}} \qquad (8)$$

where the meaning of $(\upsilon/\tau)^{1/2}$ is the relative rms of the power noise contribution.

For different values of $\tau_s/\tau_b$, $a_s$, and $a_b$ which characterize the physical conditions in the sample and for different values of the noise parameter $\upsilon$, the values of $\tau_i$ and $\tau_d$ can be optimized for maximization of the SNR. The results of such optimization are presented in Figs. 1 and 2. The SNR in the CECD scheme for a special case $\tau_s = \tau_b$ and $\upsilon = 0$ (the second condition reduces the noise to the shot noise limit) is used as a normalization factor for $SNR_p$ shown in Figs. 1a and 2a. For numerical analysis we assume that $\upsilon$ is either zero or $10^{-3}\tau_b$ (this corresponds to a relative standard deviation of about $10^{-6}$ if the signal is integrated over 1 s and $\tau_b$ is on the order of 1 ns). Two distinct regions can be identified in both panels. The region to the left from the threshold area of $\tau_s/\tau_b \approx 1 \div 2.5$ is the region where CECD provides the best SNR. In this region, $\tau_d = \tau_i = 0$ (see Figs. 1b and 2b) and the SNR is proportional to $\tau_b/\tau_s$ (as long as $\tau_b/\tau_s \ll a_b/a_s$). The region to the right from the threshold value is the region where PEGD is advantageous. In the PEGD region, $\tau_d \approx \tau_i \approx 10\tau_b$, (see Figs. 1b and 2b), the curves are practically insensitive to the values of $\upsilon$ and a maximum of the SNR is achieved at $\tau_s/\tau_b = 15 \div 30$ for a range of $a_b/a_s$ changing from below hundred to $10^6$.

A large value of $\tau_d/\tau_b$ in the PEGD region effectively eliminates the contribution from $(n_s + n_b)^2 \upsilon/\tau_i$ even if $a_b/a_s$ is large making the graphs $\upsilon$-independent. Note that in terms of the SNR, the CECD region becomes competitive only if $\tau_s/\tau_b < 10^{-2}$ (see Figs. 1a and 2a). Such a short radiative decay time (10 ps on an absolute scale as $\tau_b$ is typically $1 \div 2$ ns) is hard to achieve. The contrast of the signal in the PEGD region is on the order of $\exp(\tau_d/\tau_b) \cdot a_s/a_b$ while it is only $\tau_b/\tau_s \cdot a_s/a_b$ in the CECD region.

**Conclusion**

Interference is an effective way for eliminating spatially and/or temporarily incoherent noise which does not interfere with the auxiliary reference wave. But this method is inefficient if significant coherent noise is present on the detector. This conclusion backs the simplified analysis presented in [4]. The PEGD protocol proves to be an effective "noise eater" if its parameters are optimized. The optimization condition makes the choice between PEGD and CECD very sharp and practically discontinuous. Given a typical value of $\tau_b \approx 1 \div 2$ ns, the 25 ns luminescence lifetime of NV¯ centers in diamond nano crystals [8, 9] matches well the optimal for the PEGD scheme value of $\tau_s / \tau_b = 10 \div 30$. Therefore nano crystals of diamond activated with NV¯ centers promise to become ideal optical labels.

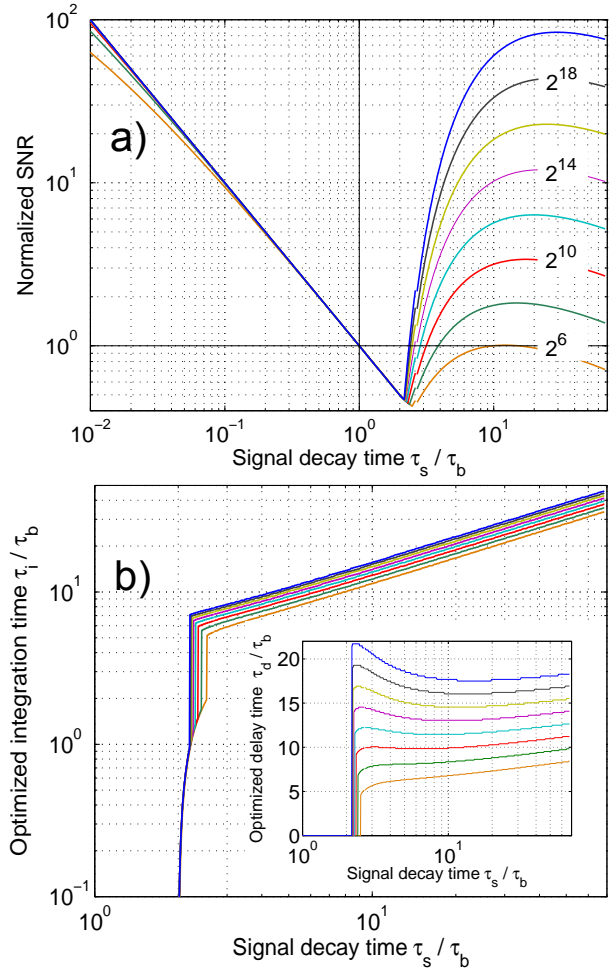

**Fig. 1.** SNR optimization when signal and background are subject to Poisson distributed shot noise. a) Highest possible SNR as a function of the signal decay time $\tau_s$. b) Most favorable for the SNR values of the integration time $\tau_i$ (main panel) and of the delay time $\tau_d$ (insert). All times are normalized to the background decay time $\tau_b$. The labels on panel a) show the relative energy ($a_b/a_s$) of the background related pulses ($a_s = 0.1$ for all curves). The vertical order of curves on panel b) is the same in panel a).

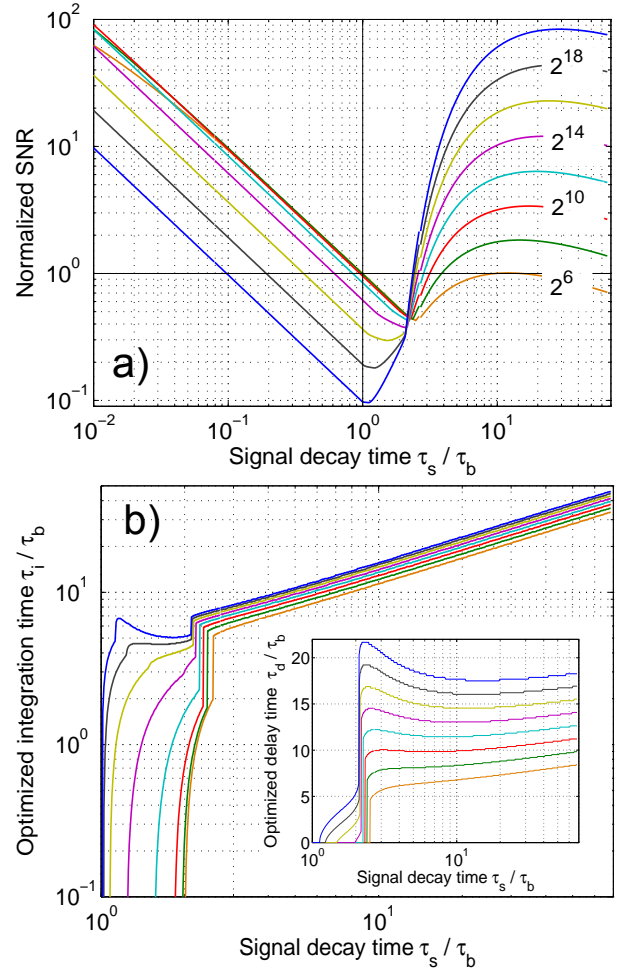

**Fig. 2.** SNR optimization when detected signal and background are subject to Poisson distributed shot noise and "white" noise with *rms* proportional to the total detected optical power. Everything else as in Fig. 1.